\documentclass[twocolumn]{aastex6}

\newcommand{\cgs}{${\rm erg}\,{\rm cm}^{-2}\,{\rm s}^{-1}$\,}

\newcommand{\lat}{\textit {Fermi}--LAT}

\slugcomment{}
\shorttitle{GeV emission from SN 2004dj}
\shortauthors{Xi et al.}

\begin{document}

\title{A serendipitous discovery of GeV gamma-ray emission from supernova 2004dj in a survey of nearby star-forming galaxies with Fermi-LAT}
\author{Shao-Qiang Xi\altaffilmark{1,5}, Ruo-Yu Liu\altaffilmark{1,5}, Xiang-Yu Wang\altaffilmark{1,5}, Rui-Zhi Yang\altaffilmark{2,6},  Qiang Yuan\altaffilmark{3}, Bing Zhang\altaffilmark{4}}
\altaffiltext{1}{School of Astronomy and Space Science, Nanjing University, Nanjing 210023, China; ryliu@nju.edu.cn; xywang@nju.edu.cn}
\altaffiltext{2}{Key Labrotory for Research in Galaxies and Cosmology, Department of Astronomy, University of Science and Technology of China, Hefei, Anhui 230026, China }
\altaffiltext{3}{Key Laboratory of Dark Matter and Space Astronomy, Purple Mountain Observatory, Chinese Academy of Sciences, Nanjing 210033, China}
\altaffiltext{4}{Department of Physics and Astronomy, University of Nevada, Las Vegas, 4505 Maryland Parkway, Las Vegas, NV 89154-4002}
\altaffiltext{5}{Key laboratory of Modern Astronomy and Astrophysics (Nanjing University), Ministry of Education, Nanjing 210023, China}
\altaffiltext{6}{School of Astronomy and Space Science, University of Science and Technology of China, Hefei, Anhui 230026, China}

\begin{abstract}
The interaction between a supernova ejecta and the circum-stellar medium drives a strong shock wave which accelerates particles (i.e., electrons and protons). The radio and X-ray emission observed after the supernova explosion constitutes the evidence of the electron acceleration. The accelerated protons are expected to produce GeV-TeV gamma-ray emission via $pp$ collisions, but the flux is usually low  since only a small fraction of the supernova kinetic energy is converted into the shock energy at the very early time. The low gamma-ray flux of the nearest supernova explosion, SN~1987A, agrees with this picture. Here we report {a serendipitous discovery of a fading GeV gamma-ray source in spatial coincidence with the second nearest supernova--SN~2004dj from our gamma-ray survey of nearby star-forming galaxies with \lat}. The total gamma-ray energy released by SN~2004dj is about $6\times10^{47}{\rm erg}$. We interpret this gamma-ray emission arising from the supernova ejecta interacting with a surrounding high-density shell, which decelerates the ejecta and converts $\sim 1$\% of the ejecta's kinetic energy to relativistic  protons. {In addition, our gamma-ray survey of nearby star-forming galaxies discovers GeV emissions from two  star-forming galaxies, i.e., Arp~299 and M33, for the first time.}
\end{abstract}


\section{Introduction}

As the supernova  ejecta expands in the circum-stellar medium, a collisionless shock  forms.  The collisionless shock  accelerates electrons to high energies, which emit photons from the radio-submillimeter through X-ray energies, as have been observed from a large number of supernovae. Protons may also be accelerated by the shocks and they can produce GeV-TeV gamma-rays via the neutral pion decay \citep{2011PhRvD..84d3003M,2011ApJ...732...58B,2015ApJ...810...63B,2019ApJ...872..157W}. Indeed, gamma-ray emission from GeV to TeV energies have been detected from a dozen of historic supernova remnants (SNRs) in our Galaxy. Although SNRs are  distinct from the early supernova as  they have already swept up a large amount of circumstellar medium (CSM), the physics governing the production of gamma-rays is the same. Thus, gamma-ray emission from  supernovae, particularly from the nearest supernova SN~1987A, has been predicted for a long while \citep{2011ApJ...732...58B,2015ApJ...810...63B}. Recently, \cite{2019arXiv190303045M} reported  a recent enhancement of the GeV emission from the SN~1987A region as observed with \lat. But the  location of this source  overlaps with several other potential gamma-ray sources, so the nature of the GeV emission remains to be clarified.

{Gamma-ray emission from Type IIn SNe and from super-luminous SNe have been searched by \cite{2015ApJ...807..169A} and \citet{2018A&A...611A..45R},  respectively.} No evidence for a signal was found, but their observational limits start to reach interesting parameter ranges expected by the theory. Recently, \cite{2018ApJ...854L..18Y} reported the detection of a variable gamma-ray source spatially and temporally consistent with a peculiar supernova, iPTF14hls. However, there is a quasar in the error circle of the \lat\/ source, which is a blazar candidate
according to the infrared data.  The lack of multi-wavelength observations of this quasar makes it difficult to conclusively
address its connection with the gamma-ray variable source.

SN~2004dj is the nearest and brightest Type IIP supernova  exploded in the galaxy NGC~2403 at a distance of about 3.5 Mpc \citep{2004IAUC.8377....1N,2004IAUC.8378....1P}.  The progenitors of SNe IIP are thought to be red-supergiants. The fast moving stellar ejecta interacts with the circumstellar medium (CSM) created by the stellar wind of the red-supergiants. SN~2004dj was detected in a wide range of wavelengths from radio, through infrared to X-rays during the first several years after the explosion \citep{2011ApJ...732..109M,2012ApJ...761..100C,2018ApJ...863..163N}.

In this paper, we report a serendipitous  discovery of a GeV gamma-ray source coincident in position with  SN~2004dj. This discovery is made when we search for GeV emission from nearby star-forming galaxies. By analysing the 11.4 years of \lat\/ data, we find a GeV source from the direction of the nearby galaxy NGC~2403. The flux of the this source, however, does not obey the well-known relation between the gamma-ray luminosity and infrared luminosity for star-forming galaxies, disfavoring the usual cosmic ray-ISM interaction origin. We also find the flux  is decaying during the \lat\/ observation period from 2008 to the present. Motivated by this, we search for transient sources in the error region of the gamma-ray emission and find that SN~2004dj lies  within the error region.

\section{Sample Selection and Data Analysis}
We perform a search for possible gamma-ray emission from  galaxies in the  IRAS Revised Bright Galaxies Sample \citep{2003AJ....126.1607S}, using 11.4 years of gamma-ray data taken by the \lat\/ telescope. This is a complete flux-limited sample of all extragalactic objects  brighter than 5.24\,Jy at 60\,$\mu m$, covering the entire sky surveyed by IRAS at Galactic latitudes $ |b|> 5^{\circ}$. In the sample, 15  infrared(IR)-bright galaxies have been detected in gamma-rays with  \lat\/ and listed in \lat\/ Fourth Source Catalog (4FGL,\citet{2019arXiv190210045T}), including six galaxies with  AGNs (i.e., Cen A, IC 4402, NGC 3067, NGC 3683, NGC 1275, NGC 3424 \citep{2019ApJ...884...91P}), seven star-forming  galaxies (SMC, LMC, M31, NGC253, M82, NGC 2146, Arp 220\citep{2010ApJ...709L.152A, 2012ApJ...755..164A, 2014ApJ...794...26T, 2016ApJ...821L..20P, 2016ApJ...823L..17G}), and two star-forming galaxies with obscured AGNs (NGC 1068, NGC 4945,\citealt{2012ApJ...755..164A}) . These galaxies are excluded from our sample for the gamma-ray emission search.

\begin{table*}[htbp]
\begin{deluxetable}{lccccccc}
\tabletypesize{\footnotesize}
\tablecaption{\label{tab:source}}
\tablewidth{0pt}
\tablehead{
\colhead{Name} &\colhead{$d_{\rm L}$}  & \colhead{Optical center} &\colhead{gamma-ray location} & \colhead{$F_{\rm 0.1-100 GeV}$}& \colhead{$\Gamma$} &\colhead{$TS$}  \\    
                           & \colhead{[Mpc]}           &  \colhead{(R.A.,Dec)}   &\colhead{(R.A.,Dec)}                      & \colhead{[$10^{-12}$\,\cgs]}        &                                      &     \\ 
             \colhead{(1)} &   \colhead{(2)} &  \colhead{(3)} &  \colhead{(4)} &  \colhead{(5)} &  \colhead{(6)} &  \colhead{(7)}
 }
\startdata
SN~2004dj & 3.49 & ($114.321^\circ, 65.599^\circ$)& ($114.395^\circ, 65.573^\circ$)$\pm$0.041$^\circ$\tablenotemark{a}& 2.29$\pm$0.51\tablenotemark{a} & 1.92$\pm$0.07\tablenotemark{a} & 73.1\tablenotemark{a}\\
Arp 299 & 47.74 & ($172.136^\circ, 58.561^\circ$) & ($172.050^\circ, 58.526^\circ$)$\pm$0.111$^\circ$ & 1.08$\pm$0.28 & 2.07$\pm$0.20 & 27.8\\
M 33 & 0.84 & ($23.475^\circ,30.669^\circ$) & ($23.609^\circ,30.784^\circ$)$\pm$0.089$^\circ$  & 1.28$\pm$0.42 & 2.33$\pm$0.24 & 25.1
\enddata
\tablenotetext{a}{These values are derived by the first 5.7 years analysis.}
\tablecomments{(1) source name;
(2) Optical Right ascension J2000;
(3) Optical Declination J2000;
(4) Best location of the $\gamma-ray$ source detected by \lat. The uncertainty of position correspond to $95\%$ containment radius, which derived by fitting the distribution of LTS to a 2D Gaussian function;
(5) 100 MeV-100 GeV $\gamma-ray$ average fluxes;
(6) Power-law spectral photon index derived by broad band spectrum fitting;
(7)$TS$ value of the gamma-ray event excess using \lat\/ data above 0.3\,GeV at the source position.}
\end{deluxetable}
\end{table*}

This work uses $\sim 11.4$ years (MET 239557417-595385929) of \lat\/ SOURCE class events with reconstructed energies between 300 MeV and 500 GeV, excluding those with a zenith angle larger than $90^\circ$ to avoid Earth limb contamination. We implement  a standard sequence of analysis steps. For each IR-bright galaxies, we select the events in a $17^\circ \times17^\circ$ region of interest (ROI) centered at the galactic center and use \textsl{gtmktime} tool to select time intervals expressed by (DATA\_QUAL $> 0$) \&\& (LAT\_CONFIG ==1). We bin the data  in 20 logarithmically spaced bins in energy and in a spatial bin of $0.025^\circ$ per pixel. We make use of recent developments of the Science Tools (v11r5p3) for likelihood analysis. The background model for each celestial ROI contains all sources listed in the 4FGL along with the standard diffuse emission background, i.e. the foreground for Galactic diffuse emission ($\rm gll\_iem\_v7.fits$) released and described by the \lat\/ collaboration through the \textsl{Fermi} Science Support Center (FSSC) \citep{2016ApJS..223...26A} and the background for spatially isotropic diffuse emission with a spectral shape described by $\rm iso\_P8R3\_SOURCE\_V2\_v01.txt$. In the likelihood analysis, we allow all sources that are separated from ROI center by less than  $6.5^\circ$ to have a free normalization to allow for the broad PSF at the lowest energies  (the $68\%$ containment radius of photons at normal incidence with an energy of 300 MeV is roughly $2.5^\circ$). This choice ensures that $99.9\%$ of the predicted gamma-ray counts is contained within the chosen radius. The normalizations of both Galactic and extragalactic diffuse emission models are left free. We employ the \textsl{gttsmap} tool to evaluate the $6^\circ \times 6^\circ$ map of the Test Statistic (TS), defined as $\rm{TS}=-2(lnL_0-lnL)$, where $L_0$ is the maximum-likelihood value for null hypothesis and $L$ is the maximum-likelihood with the additional point source with a power-law spectrum, and then survey the new gamma-ray sources when the significance of the  gamma-ray excess above backgrounds is  $TS > 25$.
The best location and uncertainty of the new  sources can be determined by maximizing the TS value and using the distribution of Localization Test Statistic (LTS), defined by twice the log of the likelihood ratio of any position with respect to the maximum. We claim a gamma-ray source associated with a target galaxy  when the target galaxy lies within the $95\%$ confidence location region of the gamma-ray source.  Note that new background sources could be found and, if so, we re-do the analysis using an updated background model including the new background sources.

\section{Results}\label{results}
Our analysis results in the detection of two new gamma-ray sources that are, respectively, spatially coincident with the galaxy Arp 299 and the galaxy M33 (see Table.~\ref{tab:source}, more details will be presented in an upcoming paper, Xi et al.~2020, in prep). In addition,  we find that the gamma-ray source reported in 4FGL (4FGL J0737.4+6535) is spatially coincident with the galaxy NGC~2403, since the best-location of 4FGL J0737.4+6535  is within the spatial extended region ($\sim 0.1^{\circ}$) of  NGC~2403. For the rest galaxies in our sample, we do not find any significant gamma-ray detection.

\begin{figure}[htbp]
\includegraphics[width=0.45\textwidth]{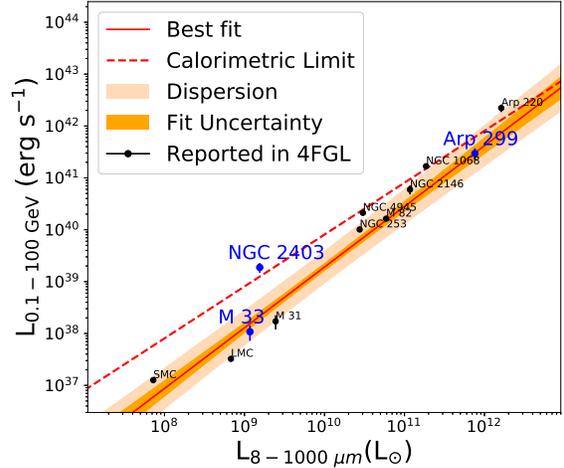}%
\caption{Gamma-ray luminosity ($0.1-100\,$GeV) vs. total IR luminosity ($8-1000$\,$\mu \rm m)$ for nearby star-forming galaxies. {The yellow band represents the empirical correlation  ${\rm log}(L_{\rm 0.1-100\,GeV}/{\rm erg\,s^{-1}})=(1.17\pm0.07){\rm log}(L_{8-1000\,\mu \rm m}/10^{10}L_{\odot})+(39.28\pm0.08)$ with an intrinsic dispersion,  which is taken to be normally distributed in the logarithmic space with a standard deviation of $\sigma_{D}=0.24$ \citep{2012ApJ...755..164A}. } The gamma-ray luminosities for the galaxies reported in 4FGL are derived from the catalog value (black points, \citealt{2019arXiv190210045T}), or based on the 11.4\,yr-averaged flux obtained in this work (blue points). The infrared luminosities are from \cite{2003AJ....126.1607S}.}
\label{fig:1}
\end{figure}

It has been found that there is an empirical correlation
between the gamma-ray luminosity in $(0.1-100)\,$GeV and total IR
luminosity ($8-1000\, \mu m$) for local group galaxies and nearby star forming galaxies \citep{2012ApJ...755..164A,2016ApJ...821L..20P}. This correlation is generally interpreted as that the gamma-ray emission arises from {cosmic-ray (CR) protons interacting with the interstellar medium (ISM) via the proton--proton ($pp$) collisions} \citep{2012ApJ...755..164A, 2019ApJ...874..173Z}.
In Figure \ref{fig:1}, we show the relation between the gamma-ray luminosity and IR luminosity for the three gamma-ray sources found in our work. While Arp 299 and M33 are well consistent with the aforementioned correlation, NGC~2403 is obviously an outlier. The latter is even above the  theoretical calorimetric limit, {which is obtained by considering} that all of the cosmic-ray energy
is converted into secondary particles and may achieve only in those gas rich starburst galaxies with luminous IR radiation (e.g., $L_{8-1000\,\mu \rm m}>10^{11}L_\odot$). This suggests that the gamma-ray emission from NGC~2403  {cannot primarily arise from the CR--ISM interaction in the galaxy.}

\begin{figure}[htbp]
\includegraphics[width=0.45\textwidth]{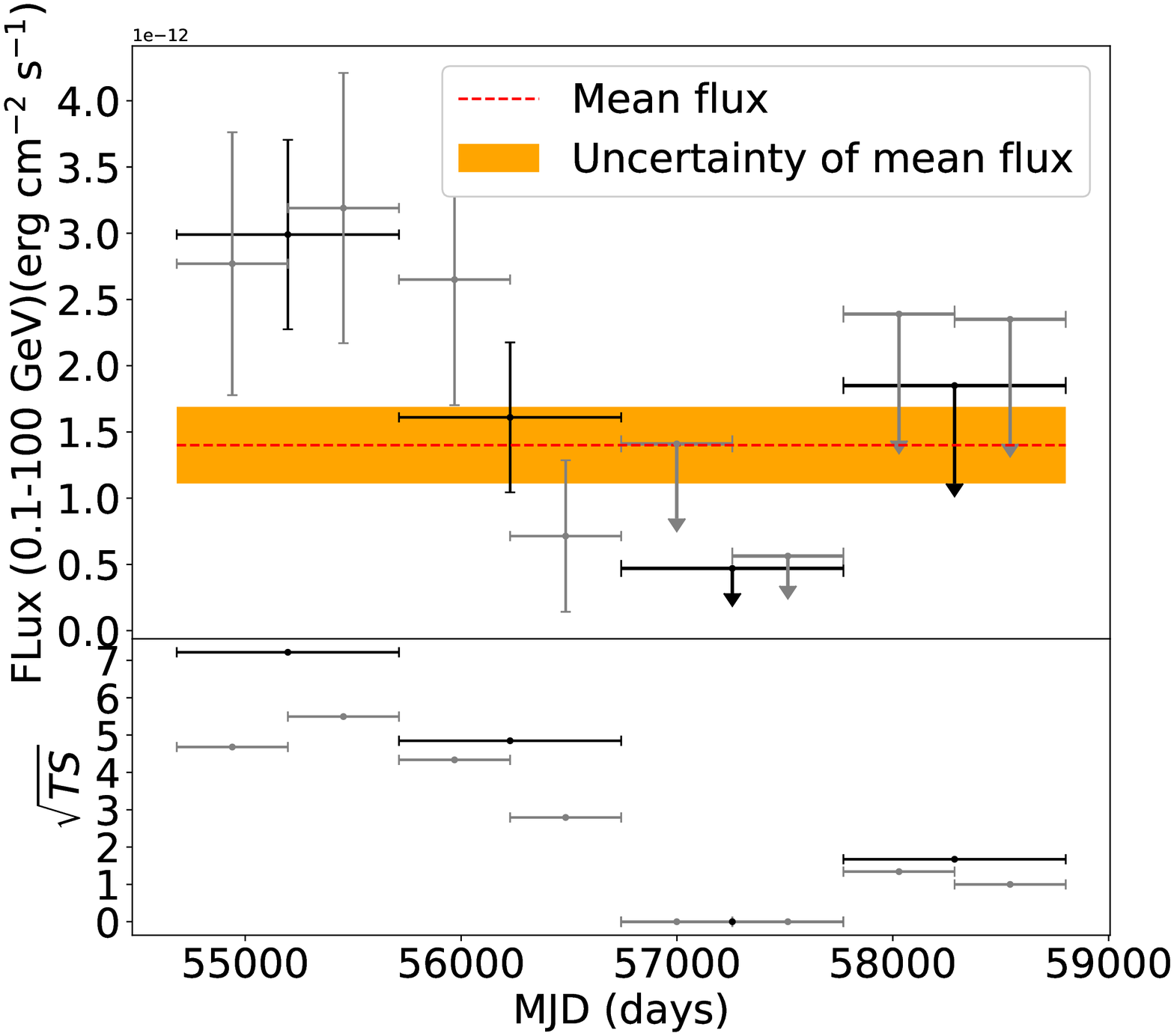}
\caption{ Light curves (upper panel) and TS values (lower panel) of the gamma-ray source 4FGL J0737.4+6535 with 4 and 8 time-bins, respectively. The mean flux is the averaged flux over the 11.4 year analysis. The upper limits at $95\%$ confidence level are derived when the TS value for the data points are lower than 4. }
\label{fig:2}
\end{figure}

\begin{figure}
\includegraphics[width=0.45\textwidth]{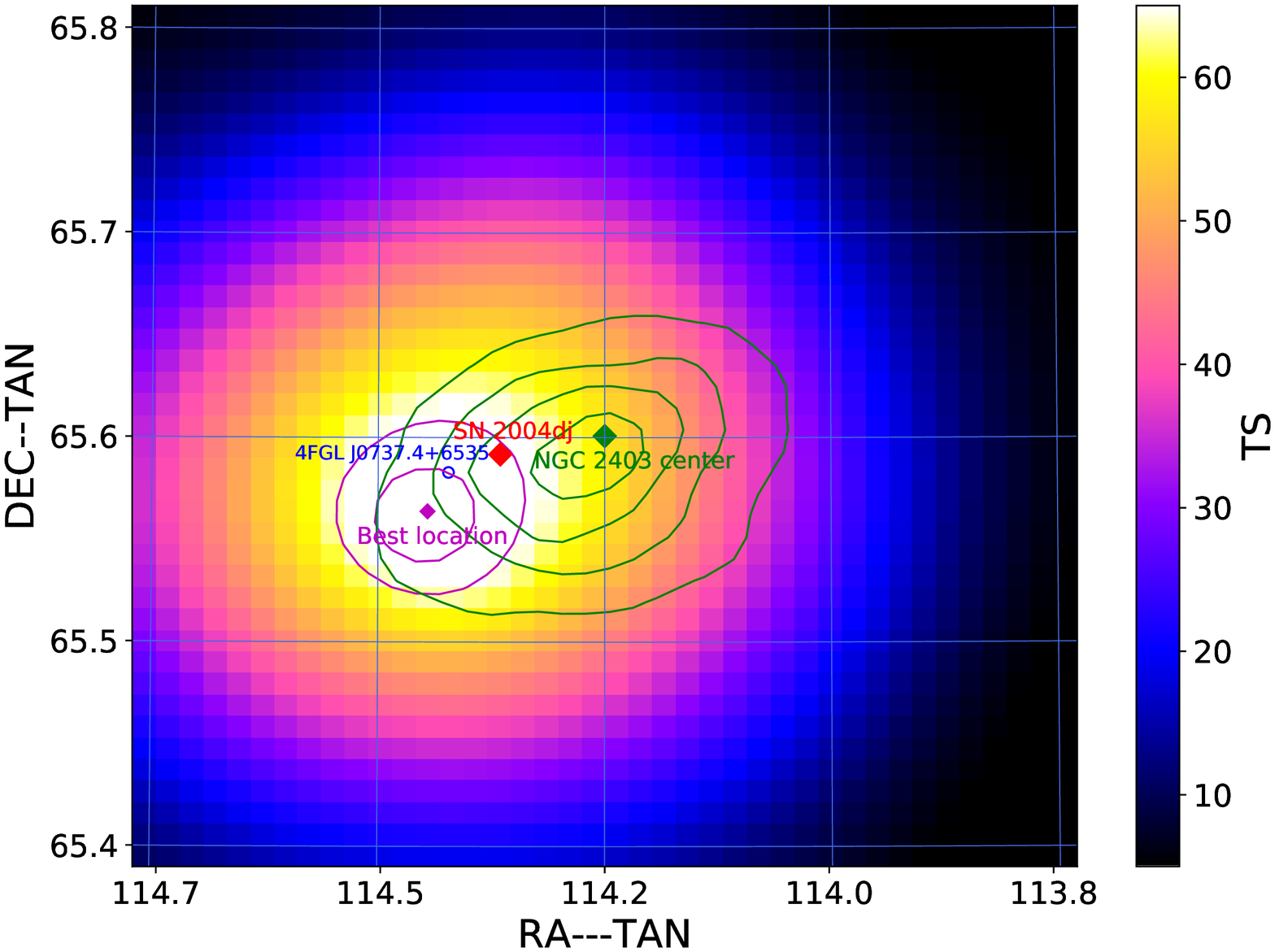}
\includegraphics[width=0.45\textwidth]{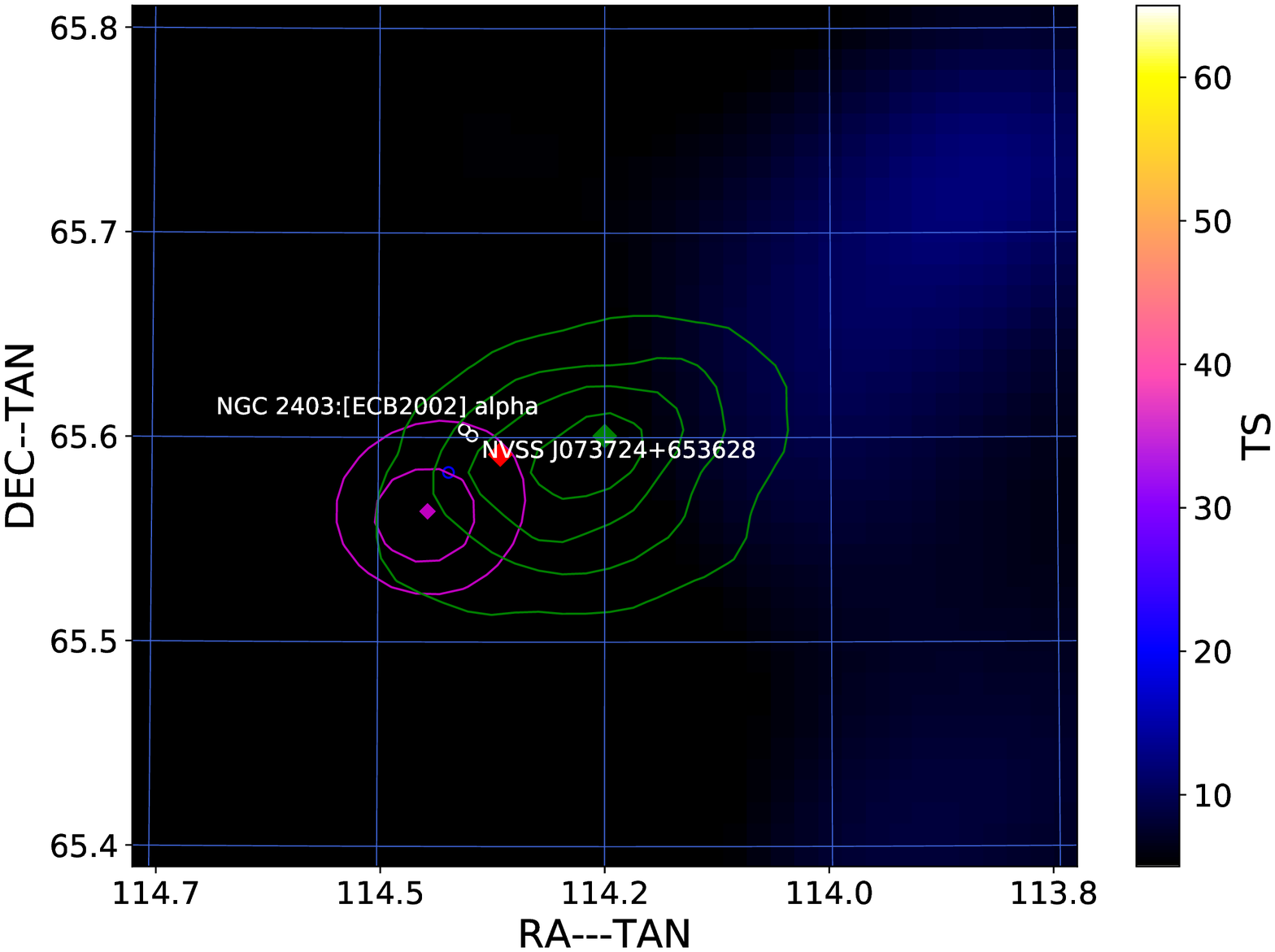}
\caption{ TS map in the energy band 0.3-500 GeV for the first 5.7 years analysis (top) and for the second 5.7 years analysis (bottom). The purple contours represent the $68\%$ and $95\%$ C.L. region of the gamma-ray source. The dark green contours represent the IR map of NGC~2403 measured by IRAS at 60$\mu$m. }
\label{fig:3}
\end{figure}

We generate the light curves of the gamma-ray source in the region of NGC~2403 with eight and four time bins respectively, which are shown in Figure \ref{fig:2}. The fluxes appear to decay with time {in both cases}. We then use  a likelihood-based statistic to test the significance of the variability. Following the definition in 2FGL \citep{2012ApJS..199...31N}, the variability index from the likelihood analysis is constructed, with a value in the null hypothesis where the source flux is constant across the full time period, and the value under the alternate hypothesis where the flux in each bin is optimized: $TS_{var} = \sum_{i=1}^{N} 2 \times (\log(L_i(F_i))-\log(L_i(F_{mean})))$, where $L_i$ is the likelihood corresponding to bin $i$, $F_i$ is the best-fit flux for bin $i$, and $F_{mean}$ is the best-fit flux for the full period assuming a constant flux.   The statistic  $TS_{var}$ is expected to be distributed, in the null case, as $\chi^2_{N-1}(TS_{var})$.  We find that the gamma-ray emission is  variable at a confidence level of $3.3\sigma$ for the analysis with 4 time bins and of $2.7\sigma$ for the analysis with 8 time bins, respectively. The variability can be also seen from Figure \ref{fig:3},  which shows  an obvious gamma-ray excess over the background in the first 5.7 years of the \lat\/ observation, but no significant excess  in the second 5.7 years. Thus, we suggest that the  gamma-ray source 4FGL J0737.4+6535 in the region of NGC~2403 is fading with time.

We use the catalogs from NASA/IPAC Extragalactic Database  (NED) and SIMBAD Database to search for  possible counterparts of the gamma-ray source. Due to the non-detection of gamma-ray emission during the second half interval of Fermi observations,  we use the data of the first 5.7 years only to re-localize the gamma-ray emission, which is shown in Fig.~\ref{fig:3}. We do not find any Galactic sources (i.e., {novae, pulsars, gamma-ray binaries}) or promising extragalactic gamma-ray emitters such as blazars located within the $95\%$ error region of the gamma-ray emission. Two faint radio sources (i.e., NVSS~J073724+653628 and NGC~2403:[ECB2002] alpha) with unknown nature are located within the $95\%$ error region. The 1.4\,GHz flux densities are $4\pm0.6 \rm mJy$ for NVSS~J073724+653628 \citep{1998AJ....115.1693C} and $\sim 1.9 \rm mJy$ for NGC~2403:[ECB2002]\,alpha \citep{2002ApJ...573..306E}. The latter one has a weak X-ray counterpart NGC~2403:[BWE2015]\,225 \citep{2015AJ....150...94B}, with a flux $\sim 3\times 10^{-16}\rm \rm ergcm^{-2}s^{-1}$ in $0.35-8\,$keV. If the radio sources are background radio galaxies, which constitute the dominant mJy radio source population at 1.4\,GHz \citep{2017NatAs...1..671N}, we can estimate the gamma-ray flux assuming that the gamma-ray emission arises from inverse Compton scattering of the radio-emitting electrons off the cosmic microwave background (CMB) photons, a reasonable expectation in light of the conclusion reached by \citet{2010Sci...328..725A}.  We use the simple scaling $F_C\approx F_S\rho_0/\rho_B$, where $F_C$ is the total Compton flux, $F_S$ is the radio flux, $\rho_0$ and $\rho_B=B^2/8\pi$ are the CMB and magnetic field energy densities, respectively. The expected gamma-ray fluxes of the two radio sources are at the level of $\sim 10^{-15}\rm erg\,cm^2\,s^{-1}$ for $B\sim 1\ \mu G$, which is about three orders of magnitude lower than the observed flux. The latter source (NGC~2403:[BWE2015]\,225) might also be an X-ray binary in NGC~2403 with an X-ray luminosity $\sim 4\times 10^{35}\,\rm erg\,s^{-1}$, but the measured gamma-ray luminosity $\sim 10^{39}\,\rm erg\,s^{-1}$ would then seem too high for it to power, since the GeV luminosities of X-ray binaries detected in our Galaxy are $\lesssim 10^{36}\,\rm erg~s^{-1}$ \citep{2013A&ARv..21...64D}. Therefore, it is unlikely that the gamma-ray emission comes from the two radio sources. 

On the other hand, we find a recent supernova explosion, SN~2004dj, occurred in the error region of the gamma-ray source. SN~2004dj is peculiar in that it is the second nearest supernova to Earth (after SN 1987A). The probability of chance coincidence between the gamma-ray source and SN 2004dj is estimated to be 0.0016 (see the Appendix.~\ref{sec:chance} for more details). Combining with the transient nature of the gamma-ray source, we suggest that it is physically associated with SN~2004dj.

\section{Discussion and Conclusion}
The GeV emission may originate from the dissipation of the kinetic energy of the SN ejecta. In this scenario, the ejecta may encounter a dense gas shell (or clump) in the vicinity of the SN progenitor and drive a strong shock. The gas shell may be formed by the stellar wind of the progenitor star \citep{1985ApJ...297..719D}. It is not clear at which epoch the shock is driven, but apparently it could be as early as a time period shortly after the SN explosion, while it should not be much later than the operation starting time of \lat, i.e. at $t_{\rm LAT,0}=1465.6\,$days (approximately 4\,yr), with respect to the SN explosion, since otherwise \lat\/ would not have collected GeV photons from the source at early time. Assuming a mass of $M_{\rm ej}=10\,M_\odot$ for the ejecta and a total kinetic energy of $E_k=10^{51}\,$ergs, the bulk of the ejecta moves with a velocity of $v_{\rm ej}=\sqrt{2E_k/M_{\rm ej}}=3000\,\rm km~s^{-1}$ and leads to a shock radius of at least $R_{\rm sh}=v_{\rm ej}t_{\rm LAT,0}\simeq 4\times 10^{16}\,$cm. The distance of the interior shell to the SN progenitor is not supposed to be larger than $R_{\rm sh}$. Given that the total gamma-ray energy detected by \lat\/ is $6\times 10^{47}\,$ergs and considering that generally $\lesssim 10\%$ of the dissipated kinetic energy could be converted to nonthermal particles, we find that $\gtrsim 1\%$ of the ejecta's kinetic energy is expected to be dissipated. This implies that a total material of $\gtrsim 0.1M_\odot$ is swept-up by the shock. As the shock sweeps through the shell, it accelerates protons and electrons therein to relativistic energies. In principle, both electrons and protons can radiate gamma rays, through inverse Compton (IC) scatterings off ambient photons or the decay of $\pi^0$ produced in $pp$ collisions, respectively.  

In the leptonic scenario, the target photon field for the IC radiation of electrons is a $\sim 500\,$K thermal radiation with a total luminosity of $L_{\rm IR}\sim 4\times 10^{38}\rm erg\,s^{-1}$ from the dust formed in the SN ejecta \citep{2011ApJ...732..109M, 2013AJ....146....2F, 2011A&A...527A..61S}, inferred from the observation of Spitzer overlapping with the first one and a half year of \lat's observation (i.e., till 2010 May 23). The photon energy density at the shock downstream is given by $u_{\rm IR}\simeq 10^{-7}(R_{\rm sh}/10^{17}{\rm cm})^{-2}\rm ergcm^{-3}$, yielding an IC cooling timescale of $t_{\rm IC}\simeq 100(E_e/50{\rm GeV})^{-1}(R_{\rm sh}/10^{17}{\rm cm})^2\,$yr. In the meantime, the accelerated electrons will also radiate in the downstream magnetic field of the shock via the synchrotron process. The ratio between the IC emissivity and the synchrotron emissivity of an electron is equal to the ratio between the energy density of the target photon field and that of the magnetic field. On the other hand, the synchrotron radiation flux should not be larger than 0.227\,mJy at 1.4GHz as observed by VLA after 1578\,days of the explosion \citep{2018ApJ...863..163N}. Therefore, as the gamma-ray flux is ascribed to the IC radiation of accelerated electrons, it constrains the downstream magnetic field to be $B<12 (R_{\rm sh}/10^{17}{\rm cm})^{-4/3}\mu$G (see Appendix.~\ref{sec:BUL} for details). Given that $R_{\rm sh}\simeq 4\times 10^{16}\rm cm$ at the operation starting time of \lat, we obtain a more conservative upper limit for the downstream magnetic field $B<41\,\mu$G, noting that it is supposed to decrease with time. This value is probably too low considering that the inferred magnetic fields of some well studied SNRs, such as Cassiopeia A, Tycho, Kepler, SN~1006 and G347.3-0.5, at much later evolution stage (i.e., from several hundred to several thousand years after the SN explosion) is $\gtrsim 0.1\,$mG \citep[][and reference therein]{2012SSRv..166..231R}, and therefore the leptonic interpretation is not favored.

\begin{figure}[htbp]
\includegraphics[width=0.45\textwidth]{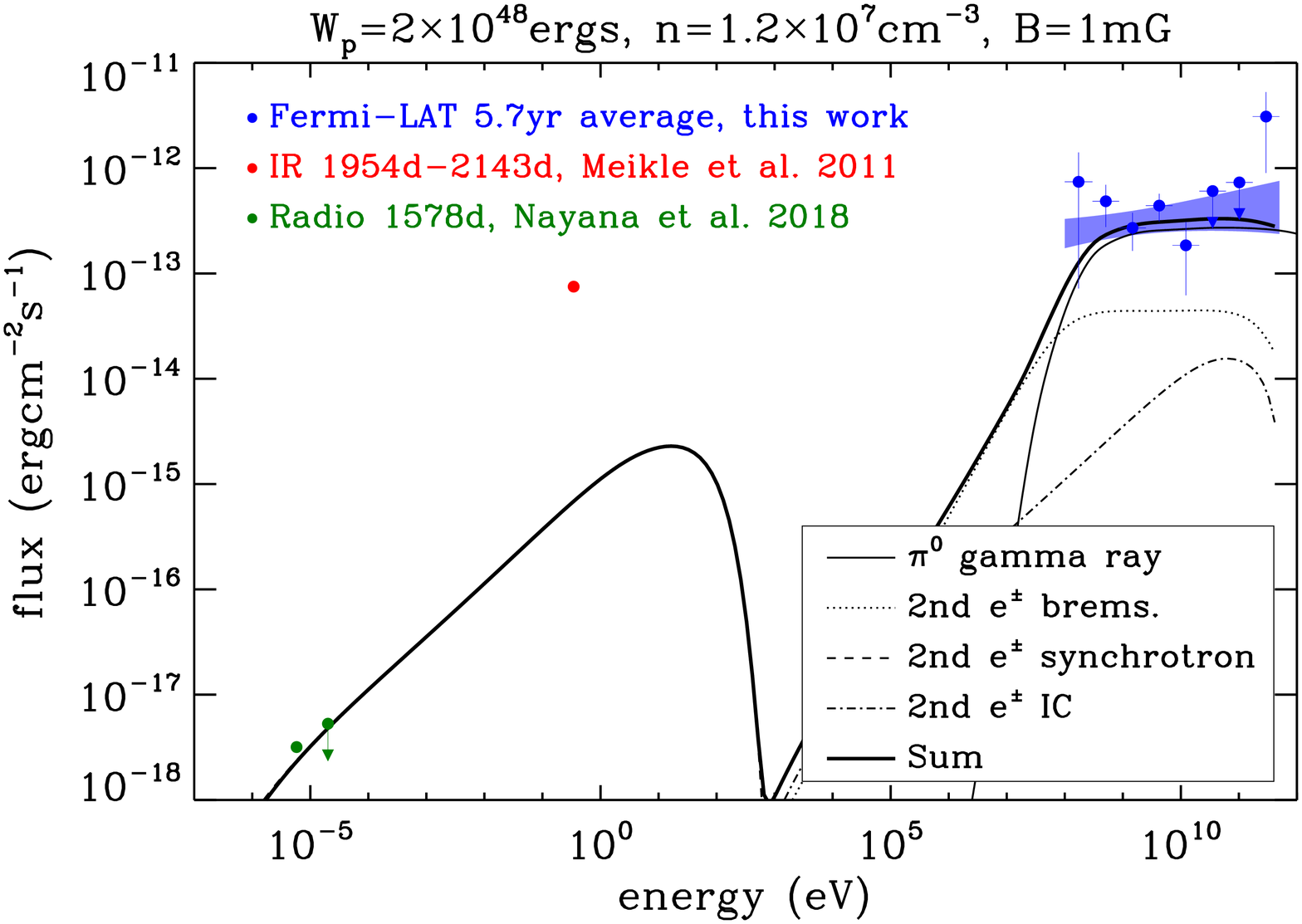}
\caption{Multi-wavelength spectrum of SN~2004dj and the theoretical expectation in the hadronic interpretation. The blue points are the first 5.7\,yr averaged flux obtained in this work. The upper limits at $95\%$ C.L. are derived when the TS value for the data points are lower than 4. The blue shaded region marks the $1\sigma$ uncertainty for a power-law fitting to the data. The red point represent the Spitzer's observation (after subtracting the contribution of the host galaxy) \citep{2011ApJ...732..109M} and the green points represent the radio fluxes measured by VLA \citep{2018ApJ...863..163N}. The thin solid curve shows pionic gamma-ray flux from $pp$ collisions. The dotted, dashed, dash-dotted curves are the flux from the bremsstrahlung, synchrotron and inverse Compton processes emitted by the secondary $e^\pm$ pairs produced in $pp$ collisions. Model parameters are shown in the top of the panel. See Section 4 for more discussion.}
\label{fig:mwsed}
\end{figure}

On the other hand, the energy loss timescale of a proton via the $pp$ collision is $t_{pp}=7\times 10^7(n/1\rm cm^{-3})^{-1}\,\rm yr,$, where $n$ is the target gas density. About $1/3$ of the lost energies will go into pionic gamma rays. We calculate the pionic gamma-ray spectrum and the secondary electron/positron pair spectrum from the $pp$ collision with the parameterized formulae given by \citet{2006ApJ...647..692K}. The result is shown in Fig.~\ref{fig:mwsed}. The total energy of the accelerated protons is assumed to be $W_p=2\times 10^{48}$ergs with a differential spectrum in a power-law distribution with slope $-2$ in the range of $1\,{\rm GeV}-1\,{\rm TeV}$. The atom density of the gas shell is set to be $n=1.2\times 10^7\rm cm^{-3}$, so that protons will cool in 5.7 years. With this assumption, the non-detection of gamma-ray emission in the second 5.7\,year could be interpreted as the cease of proton injection provided the shell's thickness $\Delta <v_{\rm ej}\times 5.7\,{\rm yr}=5.4\times 10^{16}\,$cm, while the early injected protons has already depleted their energies in $pp$ collisions. In order to suppress the synchrotron radiation of the secondary pairs below the flux observed by VLA, we find the downstream magnetic field needs to be weaker than $1\,$mG.

To summarize, using 11.4 years of \lat\/ observations, we searched for possible gamma-ray emission from nearby galaxies in the IRAS Revised Bright Galaxies Sample. In additional to the new detection of gamma-ray emission from two nearby galaxies, Arp~299 and M33, we find a gamma-ray source in the region of the galaxy NGC~2403. The gamma-ray source can not be explained as pionic gamma-ray emission from the CR--ISM interaction in NGC~2403, as it appears as an outlier of the $L_{0.1-100\,{\rm GeV}}-L_{8-1000\,\mu \rm m}$ correlation for star-forming galaxies and has a fading flux with time. On the other hand, we find that SN~2004dj, the second nearest supernova, is spatially and temporally consistent with the fading gamma-ray source. The probability of the chance coincidence of SN~2004dj with an unrelated background gamma-ray source is estimated to only 0.0016 based on the 4FGL catalog. We interpret this gamma-ray emission arising from the supernova ejecta interacting with a surrounding high-density shell, which converts $\sim 1\%$ of the ejecta's kinetic energy to relativistic protons. This shows that supernovae can accelerate cosmic-ray protons at the very early stage of their lives.

\acknowledgments

\appendix

 \section {Estimate the chance coincidence}\label{sec:chance}
\begin{figure}
\centering
\includegraphics[width=0.55\textwidth]{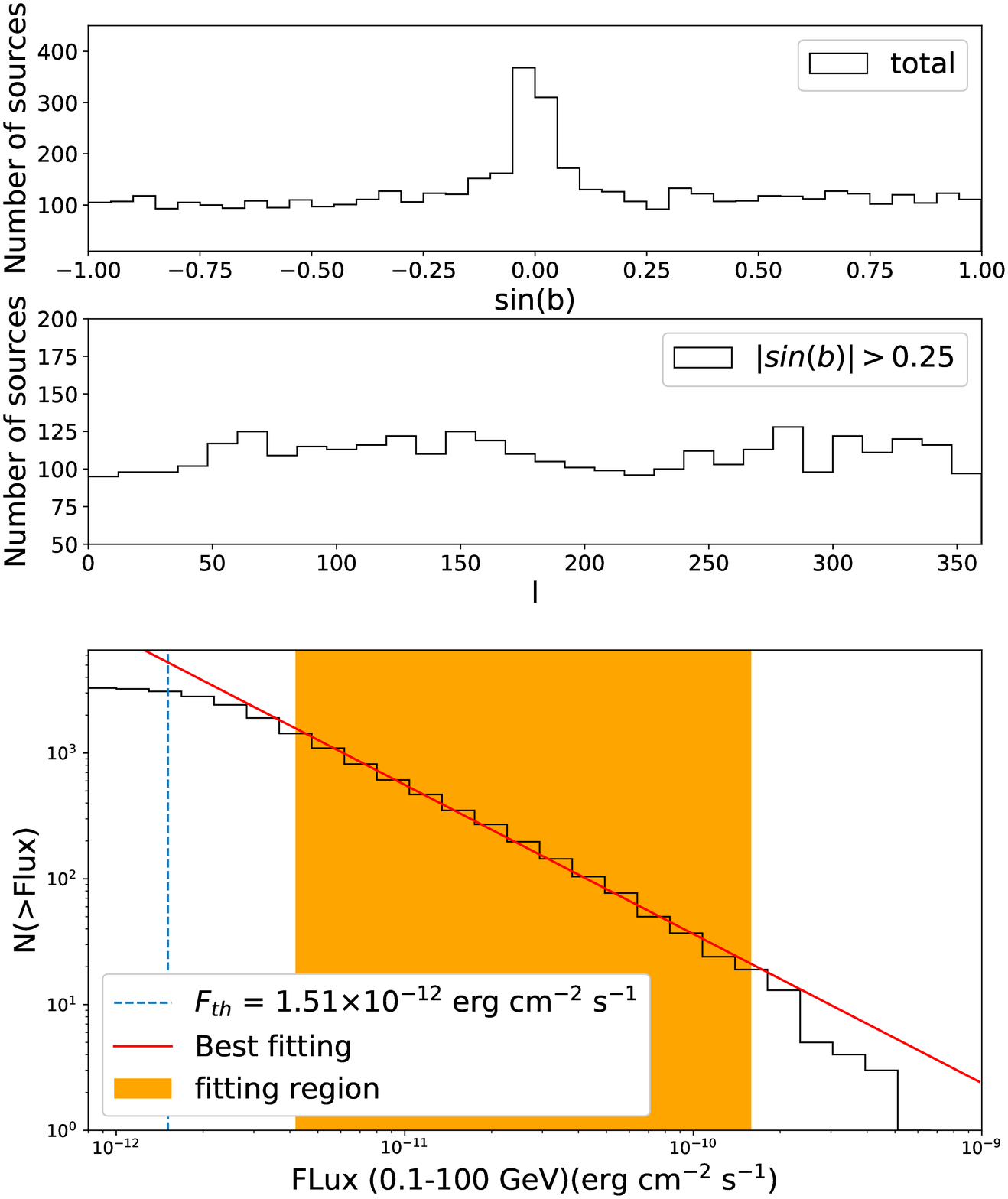}
\caption{ Upper panel: 4FGL source number distribution in Galactic latitude $b$. Middle panel: 4FGL source number distribution in Galactic longitude for $|\sin(b)|>0.25$. Lower panel: Cumulative source count distribution of 4FGL sources with $|\sin(b)|>0.25$. The blue dashed line represents the flux of 4FGL J0737.4+6535.}
\label{fig:5}
\end{figure}

Using a Poisson distribution, the chance probability of observing a background source in the position of SN~2004dj is $P_{ch}=1-\exp[-\pi (R_0^2+4\sigma_{\gamma}^2) \sum(>F_{th})]$, where $\sum(>F_{th})$ is the surface density of \lat\/ sources with fluxes higher than $F_{th}$, $\sigma_{\gamma}$ is the $68\%$ position uncertainties of \lat\/ counterpart, and $R_0$ is the angular distance between the gamma-ray location and the target source SN~2004dj.  Since the density distribution of  the 4FGL sources in the box  defined by $|sin (b)| > 0.25$ is uniform with respective  to the angle, we estimate a number density of $\sum(>F_{th})=432.5\rm\ sr^{-1}$ above the flux  $F_{th}=1.51 \times 10^{-12} \rm\ erg\,cm^{-2}\,s^{-1}$, using the power-law  fitting of the cumulative numbers of sources as a function of the threshold fluxes for the 4FGL sources, as shown in  \ref{fig:5}. Then the chance coincidence probability is estimated to be 0.0016 for $R_0=0.041^\circ$ and $\sigma_{\gamma}=0.024^\circ$.

\section{Derivation of the magnetic field upper limit}\label{sec:BUL}
{\bf Leptonic scenario:} The average gamma-ray luminosity in $0.1-100\,$GeV is $L_{\gamma, 0.1-100\,\rm GeV}=3.3\times 10^{39}\rm \,erg\,s^{-1}$. Approximating the photon index to be $-2$, we obtain the differential luminosity (i.e., $E_\gamma^2dN/dE_\gamma dt$) is $L_\gamma (E_\gamma)=L_{\gamma,0.1-100\,\rm GeV}/\log(1000)=4.8\times 10^{38}\rm erg\,s^{-1}$, independent of the photon energy. As we ascribe the gamma-ray emission to the IC radiation of accelerated electrons, $L_{\gamma}(E_\gamma)$ should be equal to the total power of electrons at energy $E_e$, where 
\begin{equation}\label{eq:E_e}
E_e=50(T_{\rm IR}/500{\rm K})^{-1/2}(E_\gamma/{1\rm GeV})^{1/2}\,\rm GeV. 
\end{equation}
The differential energy spectrum of electrons can be found by $E_e^2dN/dE_e=L_{\gamma}(E_\gamma)t_{\rm IC}(E_e)=1.5\times 10^{48}(E_e/{50\,\rm GeV})^{-1}(R_{\rm sh}/10^{17}{\rm cm})^2\,$ergs. 
According to Eq.~\ref{eq:E_e}, we find the energy of electrons responsible for $0.1\,$GeV gamma rays to be 16\,GeV. To calculate the synchrotron luminosity in radio band, we need to extrapolate the electron spectrum to energies below 16\,GeV. Note that to explain the photon index of $-2$, the electron spectral index above 16\,GeV needs to be $-3$. We assume a break at 16\,GeV and a flat spectrum ($dN/dE_e\propto E_e^{-2}$) below 16\,GeV to reduce the predicted radio flux. Such a flat spectrum is also consistent with the prediction of the canonical diffusive shock acceleration theory for strong shocks \citep{1978MNRAS.182..147B}. As such, we have $E_e^2dN/dE_e=4.7\times 10^{48}(R_{\rm sh}/10^{17}{\rm cm})^2$ for $E_e<16\,$GeV. The synchrotron cooling timescale of electrons radiating at a characteristic frequency $\nu$ is
\begin{equation}
t_{\rm syn}=4.2\times 10^7\left(\frac{\nu}{1.4\rm GHz}\right)^{-1/2}B^{-3/2}\, \rm s
\end{equation}
Then we can estimate the synchrotron luminosity by
\begin{equation}\label{eq:Lsyn_l}
L_{\rm syn}(\nu)=\frac{E_e^2dN/dE_e}{t_{\rm syn}}=1.1\times 10^{41}\left(\frac{R_{\rm sh}}{10^{17}\rm cm}\right)^2\left(\frac{\nu}{1.4\rm GHz}\right)^{1/2}B^{3/2}\, \rm erg\,s^{-1},
\end{equation}
The 1.4\,GHz luminosity observed by VLA is $4.6\times 10^{33}\rm erg\,s^{-1}$ at 1578\,days after the explosion \citep{2018ApJ...863..163N}. Since there could be other processes contributing to the flux, $L_{\rm syn}$ should be smaller than this value, and this results in $B<12(R_{\rm sh}/10^{17}\rm cm)^{-4/3}\,\mu$G. Such a strong limit on the magnetic field might in principle be relaxed, if one argues the possible existence of an extra intense radiation field around the SN during the period of \lat's detection, which in the meantime is hidden to observers, or if one argues a low-energy cutoff present in the electron spectrum above the energy responsible for radio emission. These two possibilities are, however, not supported either observationally or theoretically, so we do not further discuss them in this work.\\
  
{\bf Hadronic scenario:} In $pp$ collisions, approximately $1/3$ of produced pions are neutral pions and the other $2/3$ are charged pions which will decay into neutrinos and electrons/positrons (hereafter we do not distinguish positrons from electrons for simplicity) with a ratio 3:1. Averagely, the energy of a secondary electron is half of the energy of a gamma-ray photon produced by protons of the same energy, while the secondary electron energy production rate is also half of that of gamma rays, i.e., 
\begin{equation}
E_e^2\frac{dN_{\rm 2nd}}{dE_edt}=\frac{1}{2}L_\gamma(E_\gamma)=2.4\times 10^{38}\rm erg~s^{-1}
\end{equation}
for $50\,{\rm MeV}<E_e<50\,\rm GeV$. The radio-emitting electrons cannot be cooled during the \lat's detection period, so the accumulated secondary electron spectrum can be given by $E_e^2(dN_{\rm 2nd}/dE_e)=E_e^2(dN_{\rm 2nd}/dE_edt)\tau_{\rm GeV}$, with $\tau_{\rm GeV}$ being the duration of the GeV emission. The duration should last at least 5.7\,yr as \lat\/ detected, while it can be at longest $9.7\,$yr if the GeV emission starts shortly after the SN explosion. Similar to Eq.~(\ref{eq:Lsyn_l}), we estimate the synchrotron luminosity of secondary electrons by
\begin{equation}\label{eq:Lsyn_h}
L_{\rm syn, 2nd}(\nu)=\frac{E_e^2dN_{\rm 2nd}/dE_e}{t_{\rm syn}}=1.7\times 10^{39}\left(\frac{\tau_{\rm GeV}}{9.7\rm yr}\right)\left(\frac{\nu}{1.4\rm GHz}\right)^{1/2}B^{3/2}\, \rm erg~s^{-1},
\end{equation}
and we get $B<1.9(\tau_{\rm GeV}/9.7\rm yr)^{-2/3}\,$mG by requiring Eq.~(\ref{eq:Lsyn_h}) smaller than the measured 1.4\,GHz luminosity $4.6\times 10^{33}\rm erg\,s^{-1}$. This result is consistent with the one obtained in Section~4 (see also Fig.~\ref{fig:mwsed}) by a more accurate numerical calculation.

\end{document}